\documentclass[reprint,amsmath,amssymb,aps,showpacs,pre]{revtex4-1}
\usepackage{amsmath}    % need for subequations
\usepackage{graphicx}   % need for figures
\usepackage{verbatim}   % useful for program listings
\usepackage{color}      % use if color is used in text
\usepackage{subfigure}  % use for side-by-side figures
\usepackage{hyperref}   % use for hypertext links, including those to external documents and URLs
\usepackage{dcolumn}% Align table columns on decimal point
\usepackage{bm}% bold math

\usepackage{braket}

\begin{document}

\title{Irreversible Work and Internal Friction in a Quantum Otto Cycle of a Single Arbitrary Spin}
\author{Sel\c{c}uk \c{C}akmak$^{1,2}$}
\author{Ferdi Altintas$^{2,3}$}
\author{\"{O}zg\"{u}r E. M\"{u}stecapl{\i}o\u{g}lu$^{2,4}$}\email{omustecap@ku.edu.tr}
\affiliation{
 $^1$Department of Physics, Ondokuz May{\i}s University, Samsun 55139, Turkey\\
 $^2$Department of Physics, Ko\c{c} University, \.{I}stanbul, Sar{\i}yer 34450, Turkey\\
 $^3$Department of Physics, Abant \.{I}zzet Baysal University, Bolu 14280, Turkey \\
 $^4$School of Mathematics and Physics, Queen's University, Belfast BT7 1NN, United Kingdom}
 
\begin{abstract}
We propose an arbitrary driven spin as the working fluid of a quantum Otto cycle in the presence of internal friction. The role of total allocated time to the adiabatic branches of the cycle, generated by different control field profiles, on the extractable work and the thermal efficiency are analyzed in detail. The internal friction is characterized by the excess entropy production and quantitatively determined by studying the closeness of an actual unitary process to an infinitely long one via quantum relative entropy. It is found that the non-ideal, finite-time adiabatic transformations negatively effect the work output and the thermal efficiency of the quantum heat engine. The non-monotone dependence of the work output, thermal efficiency, entropy production and the internal friction on the total adiabatic time are elucidated. It is also found that almost frictionless adiabatic transformations with small entropy production can be obtained in a short adiabatic time. 
\end{abstract}

\pacs{05.70.Ln, 07.20.Pe, 03.65.Yz, 05.30.-d}

\maketitle

\newcommand{\dbar}{\ensuremath{\mathchar'26\mkern-12mu d}}

%\tableofcontents

\section{\label{sec:intro} Introduction}
A heat engine harvests work by manipulating energy flow between a hot energy source, a cold entropy sink and a work reservoir. The generalizations of classical heat engines to the quantum regime, i.e., quantum heat engines (QHEs), have been extensively studied, recently~\cite{scovil59,kieu04,kieu06,thomas11,altintas14,quan05,henrich07,zhang07,quan07,quan09,tonner05,turkpence16,dillen09,scully03,hardal15,gardas15,lutz14,lutz12,fialko12,zhang14,sothmann12,quan06,altintas15,geva92,rezek10,kosloff13,rezek09,thomas14,campisi15,alecce15,rezek06,wang07,feldmann00,kosloff02,feldmann03,feldmann04,allahverdyan05,feldmann06,kosloff10,wang12,feldmann12,wang13,torron13,ahn03,deffner10,wu06,plastina14,campo15,zheng15,zheng16,Ribeiro16}. The working fluid (or the working substance) of a QHE is a quantum object, while the heat baths can be classical or quantum. The cycle operation of a QHE can be governed by the quantum generalizations of some thermodynamical cycles, such as the Otto, Carnot, Brayton and Diesel cycles~\cite{quan07,quan09}, or an autonomous cycle~\cite{tonner05}. It was shown that QHEs obey the macroscopic dynamics and the classical Carnot efficiency is the limit for the efficiency of the QHEs when the heat baths are classical~\cite{quan07,quan09}. Inclusions of the quantum features of the heat baths have lead to several advantages; a QHE can extract work from a single heat bath and exceed the classical Carnot limit by exploiting quantum coherence on the hot heat bath~\cite{turkpence16,dillen09,scully03,hardal15}. Besides, there are recent statements for the universality of the Carnot limit for the case of quantum heat baths~\cite{gardas15} following the maximum entropy principle~\cite{misra15}. Several physical systems, such as a single ion~\cite{lutz14,lutz12}, ultracold atoms~\cite{fialko12}, optomechanical systems~\cite{zhang14}, quantum dots~\cite{sothmann12}, circuit and cavity quantum electrodynamic systems~\cite{quan06,altintas15} are proposed as a test bench for QHE implementations; while there are recent experimental assessments for the quantum thermal devices in a liquid-state NMR platform~\cite{batalhao14,raitz15}. Next to the fundamental explorations and the specific applications, QHEs also provide to understand the thermodynamical interpretations of the irreversibility and dissipation in the quantum domain~\cite{geva92,rezek10,kosloff13,rezek09,thomas14,campisi15,alecce15,rezek06,wang07,feldmann00,kosloff02,feldmann03,feldmann04,allahverdyan05,feldmann06,kosloff10,wang12,feldmann12,wang13,torron13,ahn03,deffner10,wu06,plastina14,campo15,zheng15,zheng16,Ribeiro16,cakmak16,acconcia15,acconcia215}.

In the present contribution, we consider an arbitrary spin interacting with a transverse time-dependent magnetic field~\cite{oliveira07,stolze08} as the working fluid of a quantum Otto engine. In the isochoric stages of the cycle, the system with a fixed Hamiltonian attains a thermal equilibrium with the heat bath, while a finite-time unitary evolution generated by a parametric time-dependent Hamiltonian develops the adiabatic stages. The main consideration is to investigate the role of quantum irreversibility and dissipation in an Otto cycle associated with the non-ideal, finite-time quantum transformations during the adiabatic stages.

The rapid change of the energy levels during an adiabatic stage causes, in general, non-adiabatic dissipation, termed as internal friction~\cite{feldmann00,kosloff13,kosloff10,rezek10}. The origin of this kind effect is fully quantum mechanical, as it arises when the system Hamiltonian at different times do not commute, i.e., $[H(t_1),H(t_2)]\neq 0$. Intuitively, friction in the adiabats can be considered as the internal resistance to the fast motion of the energy levels while compressing/expanding the working fluid. A quantum state, initially diagonal in the energy representation (such as a thermal Gibbs state), cannot follow adiabatically the time-dependent changes in the system Hamiltonian. The system state, therefore, develops coherence in the energy frame. In such a case, the energy entropy increases and an additional parasitic internal energy is stored in the working fluid. The additional energy corresponds to the waste (excess) heat which is dissipated to the heat baths during the proceeding stages of the cycle. It is found that the internal friction limits the performance of the thermal heat/refrigerator devices. Particular examples as shown recently, internal friction gradually reduces the work output, operational efficiency and power of several quantum heat engines~\cite{thomas14,wang12,wang07,alecce15}, and puts a strict restriction on the minimum temperature of the cold heat bath over absolute zero that a quantum refrigerator can extract heat from~\cite{kosloff10}.

There are possible strategies to diminish or totally eradicate the internal friction. The first way relies on the quantum adiabatic theorem and provides a frictionless adiabatic transformation. The system state follows the instantaneous energy eigenstates, which entails infinitely long times, being, however, in the expense of finite power output, and may open the system to decoherence. The strategies named as "quantum lubrication"~\cite{feldmann06,rezek10} and "shortcuts to adiabaticity"~\cite{campo15,acconcia15} are the minimizing strategies for the destructive frictional effects. The former one~\cite{feldmann06,rezek10} adds external noise on the control parameters which suppresses the frictional effects, resulting, however, imperfect control of the external parameters during the adiabatic stages. The latter one~\cite{campo15,acconcia15} requires a successful design of a control pulse where at the end of the adiabatic stage the frictional effects are minimized.

The role of internal friction in thermal devices has been recently analyzed in an {\it ad hoc} manner, in which the source of friction is not definite and it is added to the system phenomenologically~\cite{wang07,feldmann00}. In our setting, its source is explicit and it is caused by the misalignment between the external transverse and internal longitudinal magnetic fields~\cite{rezek10,kosloff13,thomas14,alecce15,rezek06}. Our main interest is two-fold. On one hand, we analyze how the work output and the thermal efficiency are effected by the internal friction. On the other hand, we quantitatively determine the internal friction and its explicit role in the cycle. We consider four different shaped control fields that  generate the adiabatic transformation. Since each driving pulse defines different rate for the transformation of the system Hamiltonian, they are found to lead to pronounced effects on the performance characteristic of the Otto cycle. Our results reveal that considering different possibilities for a driving pulse allows us to improve the performance of the quantum Otto engine, to explore new features of the quantum friction, and to find a solution for the almost frictionless cases in finite time adiabatic transformations. Our considered scheme uses an elementary system (an arbitrary single spin) which is easy to implement in NMR setups~\cite{batalhao14,raitz15} and clearly identifies and fights against the fundamental source of friction (non-commuting free and control parts of the Hamiltonian). 

\section{\label{sec:theory} Theory}
\subsection{\label{sec:worksubs}Working Fluid}
We consider a single spin in a time-dependent magnetic field as the working fluid of a quantum Otto engine. The adiabatic branches of the cycle are assumed to be generated by a time-dependent Hamiltonian of the form
\begin{equation}\label{eq:ssh}
H(t)=B_0 I_z + B(t) I_x.
\end{equation}
Here $B_0$ is the static magnetic field along the $z$-axis, while $B(t)$ is the time-dependent magnetic field along the $x$-axis and changes the system Hamiltonian in the adiabats. $I_z$ and $I_x$ are the components of the spin angular momentum which obey the standard commutation $[I_x,I_z]=-iI_y$. Here we restrict ourselves to the spin-$I$ values, $I=1/2,1,3/2,2$. Throughout the paper, we set $\hbar=1$ and the gyromagnetic ratio $\gamma=1$. One should remark that our choice of model Hamiltonian is not arbitrary. We want to examine the simplest elementary model, containing the fundamental cause of friction, non-commuting free and control terms in the Hamiltonian. Moreover, the Hamiltonian~\eqref{eq:ssh} is in the generic form commonly used in liquid-state NMR experiments of a single nuclei~\cite{oliveira07,stolze08}.

\subsection{\label{sec:qoc} Quantum Otto Cycle}
We will characterize the quantum Otto cycle by investigating the extractable work and its thermal efficiency. Quantum Otto cycle consists of two quantum isochoric and two adiabatic processes. In the adiabatic stages, the working fluid does not exchange any heat with the surroundings and performs only work, while in the isochoric stages, only heat is transferred. The work performed or the heat exchanged in the strokes of the cycle can be, therefore, formulated through the change in the internal energy. According to quantum thermodynamical version of the first law of thermodynamics~\cite{quan07}, the change in the internal energy is $dU=\dbar Q-\dbar W$, where $U=Tr(\rho H)$ is the internal energy, $Q$ is the heat added to the system and $W$ is the work performed by the working fluid. The details and the formulation of the cycle are as follows.

\textit{Isochoric Heating:} The working fluid with Hamiltonian $H^{(1)}=B_0 I_z+B_1 I_x$ is in contact with a hot energy source at temperature $T=T_1$. The density matrix of the working fluid at the end of the stage can be given as $\rho_T^{(1)}=\exp(-\beta_1 H^{(1)})/Z_1$, where $Z_1=Tr[\exp(-\beta_1 H^{(1)})]$ and $\beta_1 = 1/T_1$ (in units of $k_B=1$).

\textit{Adiabatic Expansion:} The working fluid is decoupled from the hot heat bath and undergoes finite-time adiabatic process. The Hamiltonian  $H^{(1)}$ at $t=0$ is changed to $H^{(2)}=B_0 I_z+B_2 I_x$ at $t=\tau/2$. The time evolution of the density matrix can be governed by Liouville-von Neumann equation $\dot{\rho}(t)=-i[H(t),\rho(t)]$, where $H(t)$ is given in Eq.~\eqref{eq:ssh} and the initial condition is $\rho(t=0)=\rho_T^{(1)}$. The adiabatic transformation can be obtained with different shaped control fields, for instance, in the form as $B(t)=B_1+(B_2-B_1)\sin(\pi t/\tau)$ or $B(t)=B_1+(B_2-B_1)(2t/\tau)^n$ with $n=1/2,1,2$. Here $B(t=0)=B_1$ and $B(t=\tau/2)=B_2$ in each case. The work performed by the working fluid in the process can be written as $W_{I}=Tr[H^{(1)} \rho_T^{(1)}]-Tr[H^{(2)} \rho^{(2)}]$, where $\rho^{(2)}=\rho(t=\tau/2)$ is the final density matrix in the unitary evolution.

The unitary evolution represents a closed system, so the evolution is adiabatic in the sense of the fact that there would be no heat exchange with the surrounding during the process. On the other hand, the quantum adiabatic theorem is not expected to hold in the system due to the finite-time nature of the adiabatic evolution and the non-commutativity of the free and control terms in the Hamiltonian~(\ref{eq:ssh}). As a result, a phenomenon termed as internal (quantum) friction arises~\cite{feldmann00,kosloff13,kosloff10,rezek10}. As we will show, this has profound influences on the performance characteristic of the Otto cycle.

\textit{Isochoric Cooling:} The working fluid is in contact to a cold entropy sink at temperature $T=T_2$. The density matrix at the end of the stage would be $\rho_T^{(2)}=\exp(-\beta_2 H^{(2)})/Z_2$, where $Z_2=Tr[\exp(-\beta_2 H^{(2)})]$, $\beta_2=1/T_2$ and the Hamiltonian $H^{(2)}=B_0 I_z+B_2 I_x$ is fixed during the stage. The amount of released heat can be calculated as $Q_2=Tr[H^{(2)}(\rho_T^{(2)}-\rho^{(2)})]$, where $\rho^{(2)}$ is the final density matrix of the adiabatic expansion.

\textit{Adiabatic Compression:} The time evolution of the process can be given by $\dot{\rho}(t)=-i[H(t),\rho(t)]$, where $\rho(t=0)=\rho_T^{(2)}$ and $H(t)$ is given in Eq.~\eqref{eq:ssh} with $B(t)=B_2+(B_1-B_2)\sin(\pi t/\tau)$ or $B(t)=B_2+(B_1-B_2)(2t/\tau)^n$ ($n=1/2,1,2$). The Hamiltonian $H^{(2)}$ at $t=0$ is transformed back to $H^{(1)}$ at $t=\tau/2$. The work performed by the fluid can be given as $W_{II}=Tr[H^{(2)} \rho_T^{(2)}]-Tr[H^{(1)} \rho^{(1)}]$, where $\rho^{(1)}=\rho(t=\tau/2)$ is the final density matrix in the evolution.

Bringing the system back in contact with the hot heat bath will return the system to its initial state. The amount of absorbed heat from the hot heat bath can be then calculated as $Q_1=Tr[H^{(1)}(\rho_T^{(1)}-\rho^{(1)})]$. The change of internal energy is zero in the cycle. Therefore, the net extractable work can be given as $W=W_{I}+W_{II}=Q_1+Q_2$ with thermal efficiency, $\eta=W/Q_1=1+Q_2/Q_1$. 

We should stress here that our above approach relies on two presumptions. We have first assumed the perfect thermalization of the system at the heat bath temperature at the end of  the isochores. Indeed, the proper choice of the Lindbladian produces the Boltzmann distribution~\cite{oliveira07,stolze08}. It can also be shown to be realized in NMR setups~\cite{batalhao14,raitz15,oliveira07,stolze08}. Second, our proposed smooth field-change protocols ensure non-crossings between the adiabatic energy levels. It was shown recently that if level crossing occurs, quasi-static reversible changes are not the optimal processes~\cite{allahverdyan05}.

\subsection{\label{sec:inrfricadi} Entropy Production in Adiabatic Branches}
In present study, we will analyze the role of the total allocated time $\tau$ to the adiabatic branches of the Otto cycle on the extracted work and the thermal efficiency. A reversible quasi-static transformation of the system Hamiltonian (in the sense of quantum adiabatic theorem) ensures the perfect follow of the system eigenstates without changing the initial level populations. On the other hand, finite time stages, in general, cannot accomplish perfect adiabatic transformation; the system state deviates from equilibrium and produces coherence in the energy frame. The coherence generation can be associated with the entropy production, signaling the presence of internal friction in the system. In principle, such effects arise when the Hamiltonian at different times does not commute. Using Eq.~\eqref{eq:ssh}, one can easily show that the commutator is always different than zero, since the above proposed magnetic fields are non-uniform ($B(t_1)\neq B(t_2)$),  i.e., $[H(t_1),H(t_2)]=-iB_0\left(B(t_1)-B(t_2)\right)I_y$.

The deviation from the perfect adiabaticity can be characterized by analyzing the difference between von Neumann entropy $S_V=-Tr[\rho \ln \rho]$ and the energy based (Shannon) entropy $S_E=-\sum_{i} p_i \ln p_i$, where $p_i=Tr[\left|i\right\rangle\left\langle i\right|\rho]$ are the occupation probabilities of the energy levels. For the quantum states where $\rho$ is diagonal in the energy frame (such as the thermal states at the beginning of the adiabats and under quantum adiabatic theorem), we have $S_E=S_V$. In addition, the von Neumann entropy does not change in the adiabatic branches, since the adiabats are formulated through a unitary transformation and $S_V$ remains invariant under unitary transformations. On the other hand, the finite-time transformations can redistribute the level populations due to the coherence generation, consequently $S_E$ will be different than $S_V$. The von Neumann entropy is always a lower bound to $S_E$, i.e., $S_E \geq S_V$. Therefore, the deviation from the perfect adiabaticity can be characterized by the increase in $S_E$, representing the signatures of the internal friction. The sum of the increments in the energy entropy according to the end points of two adiabatic branches will be used, at first hand, for the characterization of the internal friction in our setting. It will be denoted as $\Delta S_E$. We refer the reader to Refs.~\cite{rezek06,kosloff13} for more discussions on the topic.

\section{\label{sec:results} Results}

\subsection{\label{sec:workeff} Work and Efficiency}

\begin{figure}
\begin{center}
\includegraphics[scale=0.21]{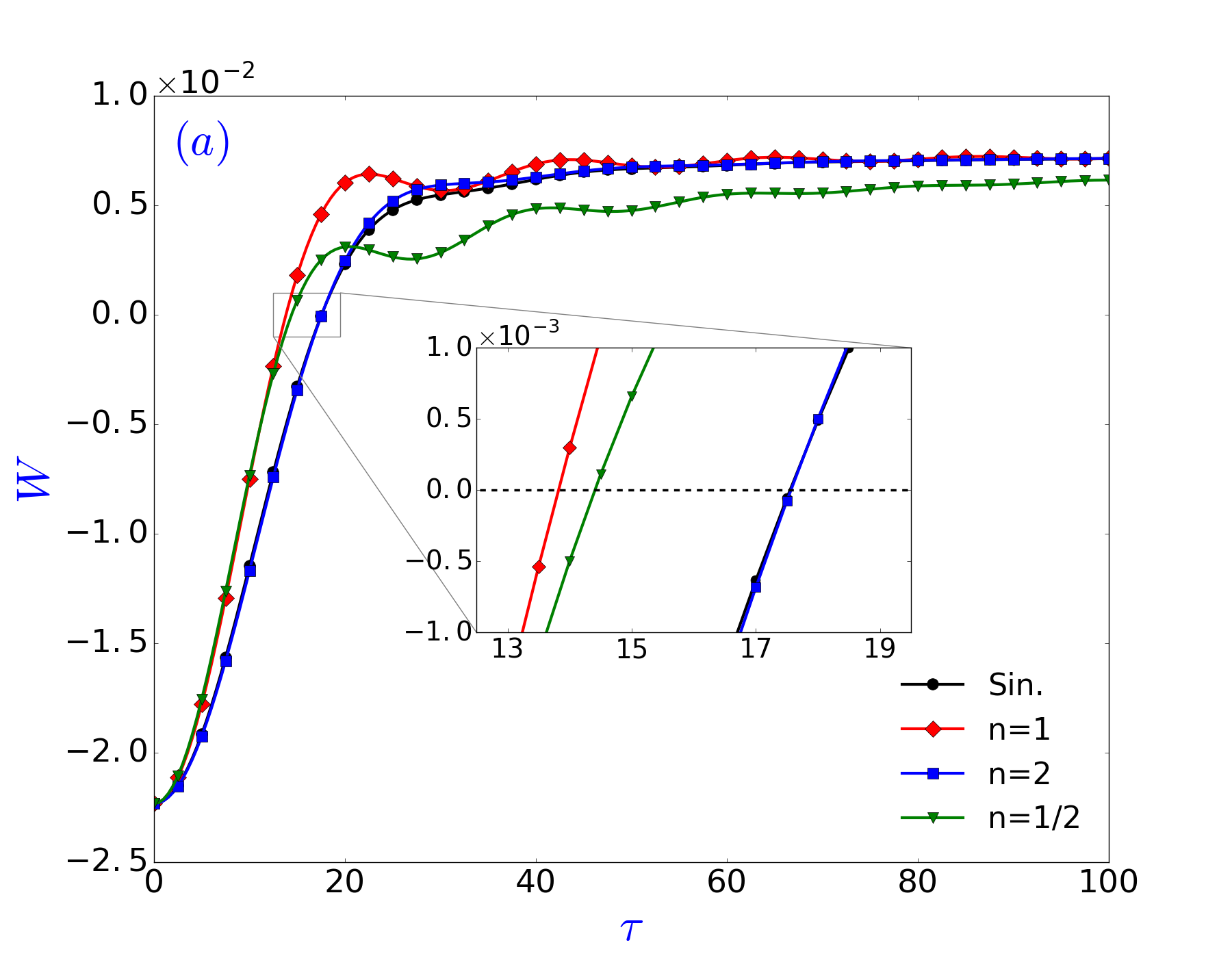}
\includegraphics[scale=0.21]{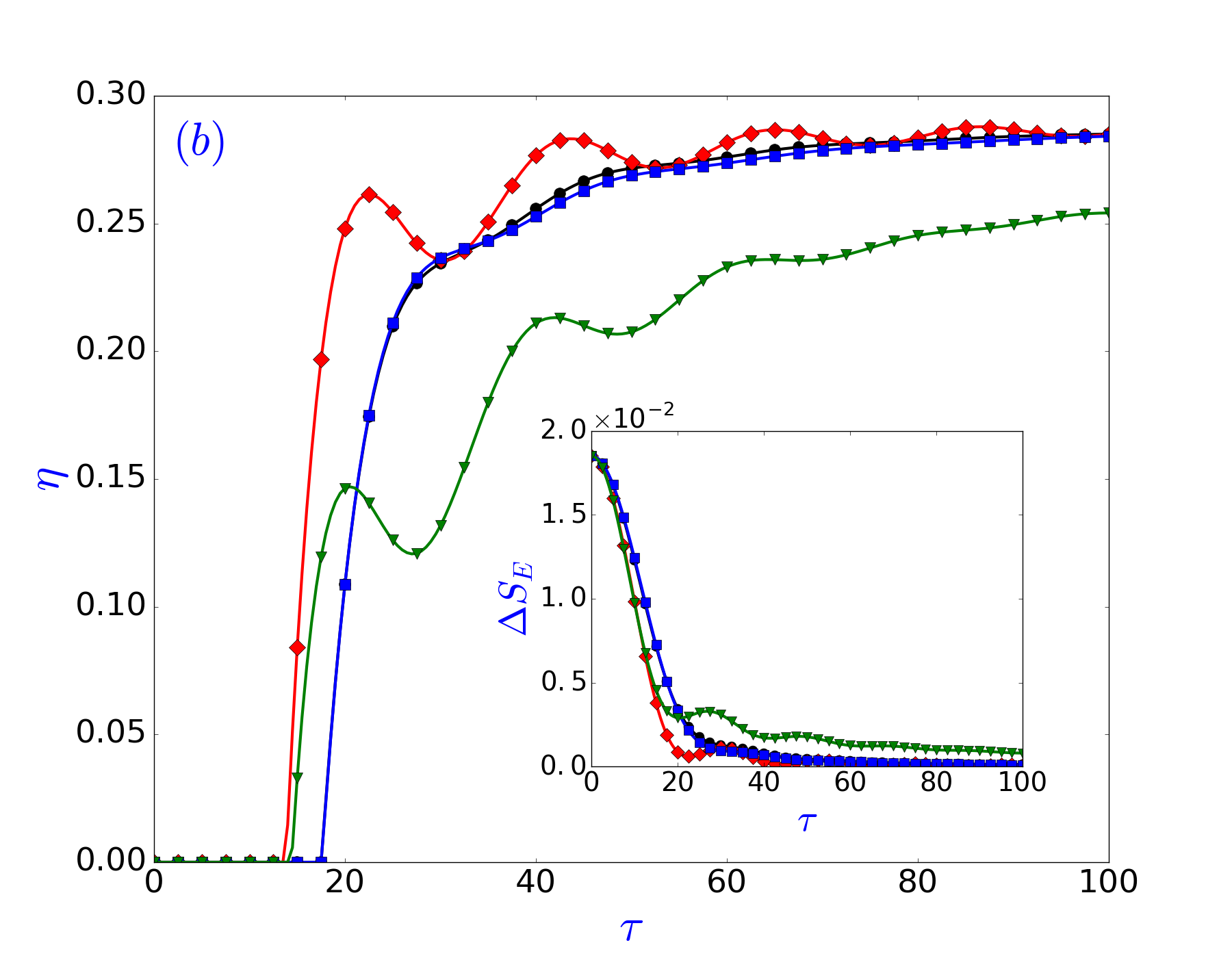}
\caption{\label{fig:0}(Color online.) (a) Work obtained in a cycle and (b) thermal efficiency versus total adiabatic time $\tau$ for the spin-$1/2$ and the parameters $B_0=B_1=0.5$, $B_2=0.05$, $T_1=2$ and $T_2=1$. The adiabatic stages are generated by four different driven pulses as discussed in the text. The inset in (b) shows the total entropy production in the adiabats versus $\tau$. The inset in (a) shows a magnification for $W=0$ region.}
\end{center}
\end{figure}

Now we investigate the effect of total allocated time $\tau$ to the adiabats, generated by different driven pulses, on the work output and the thermal efficiency. In Fig.~\ref{fig:0}, we plot the work obtained in a cycle and the thermal efficiency as a function of $\tau$ for a spin-1/2 working fluid. Analyzing $\tau$ dependence of the performance parameters of the Otto cycle demonstrate pronounced effects. There is a strong entropy production in the adiabats (inset of Fig.~\ref{fig:0}(b)), showing the presence of internal friction in the system. One can define a critical time $\tau_c$ for the adiabats. When the adiabatic stages last below this critical time, $\tau<\tau_c$, the engine cannot harvest positive work, while when $\tau>\tau_c$, we have $W>0$. As shown in the inset of Fig.~\ref{fig:0}(a), this critical time depends on the type of the control field. We should stress here that for the times, where $W<0$, the engine operation is not a refrigerator. Here heat is absorbed from the hot heat bath, while some amount is rejected to the cold one, but $|Q_2|>Q_1>0$ so $W<0$. One can note here that the qubit Otto engine cannot accomplish the frictional effects when the adiabatic branches are allotted shorter time intervals than $\tau_c$. The total entropy production $\Delta S_E$ is high in the negative work region compared to the positive one. We further studied the mutual relation between the work output and the total entropy production. We obtain reciprocal linear plots (not shown here) which coincide for each considered pulses. This is important because the reduction of the work output and the entropy production are not only the hallmarks of the internal friction, but also its quantitative measure. More results, discussions and the quantitative analysis for the internal friction  will be given in the following text.

Analyzing the effects of the total allocated time $\tau$ to the adiabatic branches on $W$ and $\eta$ contracts with the common belief. It is generally expected that $W$ and $\eta$ should depend on the time $\tau$ in a linear manner up to the reaching their maximums (quantum adiabatic theorem). Indeed, the monotonic dependence of the extracted work on $\tau$ is confirmed in recent studies~\cite{thomas14,feldmann00,alecce15}.  As shown in Fig.~\ref{fig:0}, the adiabatic branches generated by the sinusoidal or square ($n=2$) pulse almost justify this expectation. On the other hand, for the linear ($n=1$) and square root ($n=1/2$) pulses, $W$ and $\eta$ show oscillatory behavior as a function of $\tau$. This means that the Otto engine can produce higher work or harvest work with higher efficiency for shorter adiabatic times compared to the longer ones. For instance, at $\tau \sim 20$, we have $\eta \sim 0.26$, while at $\tau \sim 30$, we have $\eta \sim 0.22$ for the linear pulse ($n=1$). Similar examples can be given for the work output and the square root ($n=1/2$) pulse cases. It is interesting to note that for the time intervals where $W<0$, the work output and the entropy production are monotonic functions of $\tau$ for each pulse cases. The non-monotone behavior emerges just after the engine can harvest positive work, $W>0$. We should stress here that the non-monotonic dependence of work and efficiency on the driving time and also the role of asymmetric adiabatic strokes on the optimization of the Otto cycle for a harmonic oscillator working substance have been discussed in a recent paper~\cite{zheng16}.

We can conjecture that two extreme values of $\tau$ for the adiabatic branches determine the allowed range of the work output~\cite{thomas14,feldmann12}. In the case, where zero allocation time interval to the adiabats is allowed (i.e., $\tau\rightarrow 0$), the lower bound of the work output can be obtained. This corresponds to the largest entropy production in the adiabats. Even if the system Hamiltonian is changed in an extremely fast process, the density matrix remains invariant; $\rho_T^{(1)} \rightarrow \rho_T^{(1)}$ and $\rho_T^{(2)} \rightarrow \rho_T^{(2)}$ in the adiabatic expansion and compression stages, respectively. Using the above formulation, one can determine the lower bound of the work output for the given parameters in Fig.~\ref{fig:0} as $W_{\text{lb}}/k_B T_2 \cong -2.233 \times 10^{-2}$.

On the other hand, the maximum amount of positive work is extracted when the adiabatic stages proceed quasi-statically (i.e., $\tau\rightarrow \infty$). Internal friction in the system is completely eliminated, so there would be no increase in the entropy. The quantum adiabatic theorem holds; the occupation probabilities of each eigenstates are maintained in an infinitely slow process. The density matrix follows the instantaneous eigenstates of the Hamiltonian. One can then determine the upper bound of the extractable work for the given parameters in Fig.~\ref{fig:0} under the quantum adiabatic theorem~\cite{quan07} as $W_{\text{up}}/k_B T_2 \cong 7.277 \times 10^{-3}$. For the plots in Fig.~\ref{fig:0}, we have numerically verified that the work output lies in the range $W_{\text{lb}} < W < W_{\text{up}}$.

We should stress here that a vast of studies in the literature relies on the validity of the quantum adiabatic theorem~\cite{kieu04,kieu06,thomas11,altintas14,quan05,henrich07,zhang07,quan07,quan09}. However, it entails very low power output and may open the system to decoherence as it may require timescales much longer than the internal timescale of the working fluid. Under quantum adiabatic theorem, our system becomes equivalent to the Kieu's model of a qubit Otto engine with an energy gap $\delta_i=\sqrt{B_0^2+B_i^2}$ ($i=1,2$ for the hot and cold isochores, respectively)~\cite{kieu04,kieu06}. The thermal efficiency of the cycle can be given as~\cite{kieu04,kieu06}: 
\begin{equation}\label{eq:eff}
\eta_{m}=1-\frac{\delta_2}{\delta_1}.
\end{equation}
The positive work condition $(W>0)$ of the engine is $T_1>\delta_1/\delta_2T_2$~\cite{kieu04,kieu06}, consequently $\eta_m$ is bounded above by the classical Carnot efficiency ($\eta_c=1-T_2/T_1$). Our numerical results in Fig.~\ref{fig:0} show that the maximum thermal efficiency is obtained under quantum adiabatic theorem, that is $\eta\leq\eta_m\cong 0.289$. It is therefore our proposed Otto engine can harvest highest work at highest efficiency under quantum adiabatic theorem. That is to say quasi-static stages are the optimal processes~\cite{allahverdyan05}. It is important to note that almost frictionless solutions with very small entropy production can also be obtained for finite times of $\tau$; specifically the adiabatic evolutions generated by the linear driven pulse ($n=1$) provides almost frictionless solutions at smaller $\tau$ ($\tau>40$), while $n=1/2$ pulse requires much longer times, ($\tau>600$).

\begin{figure}
\begin{center}
\includegraphics[scale=0.21]{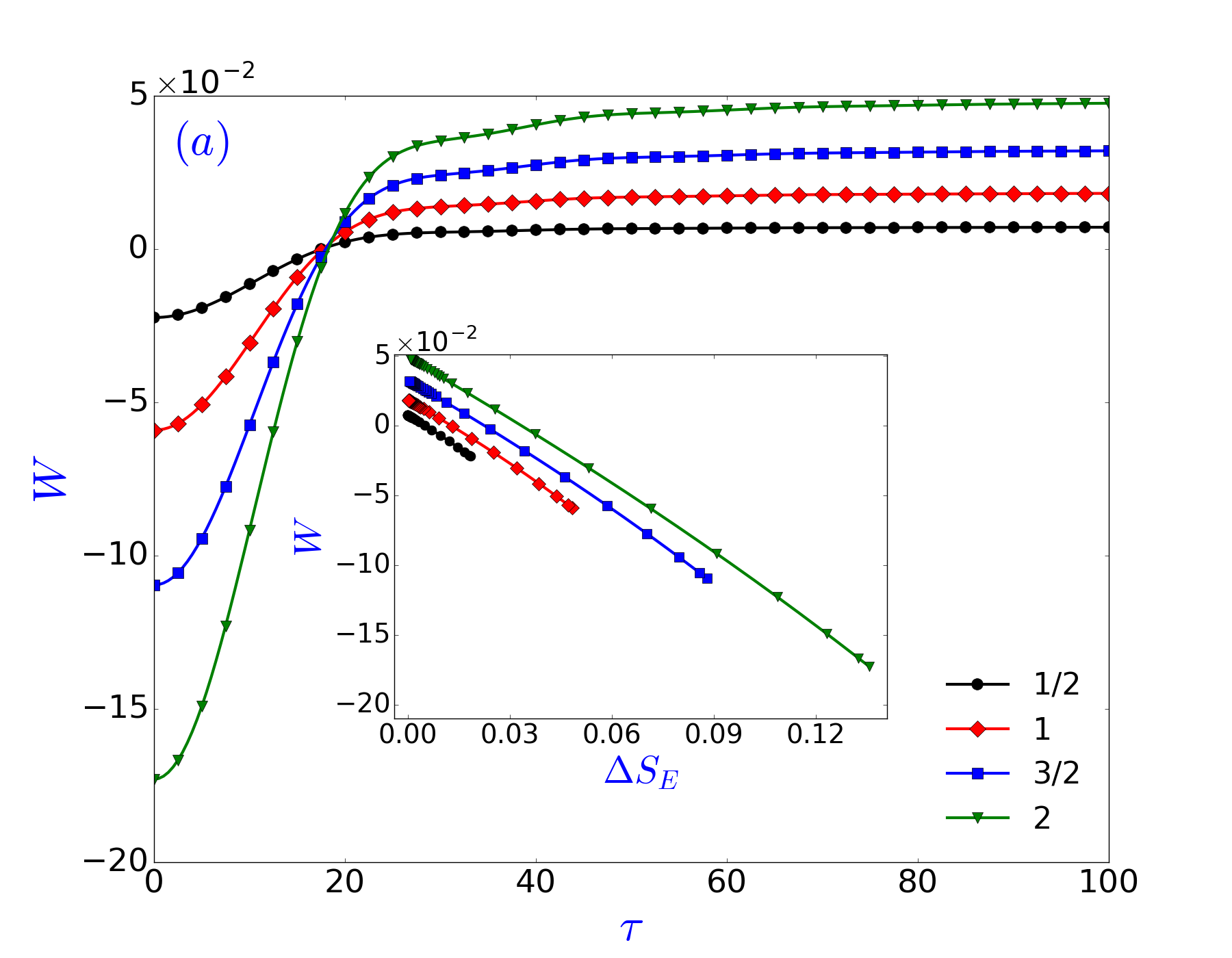}
\includegraphics[scale=0.21]{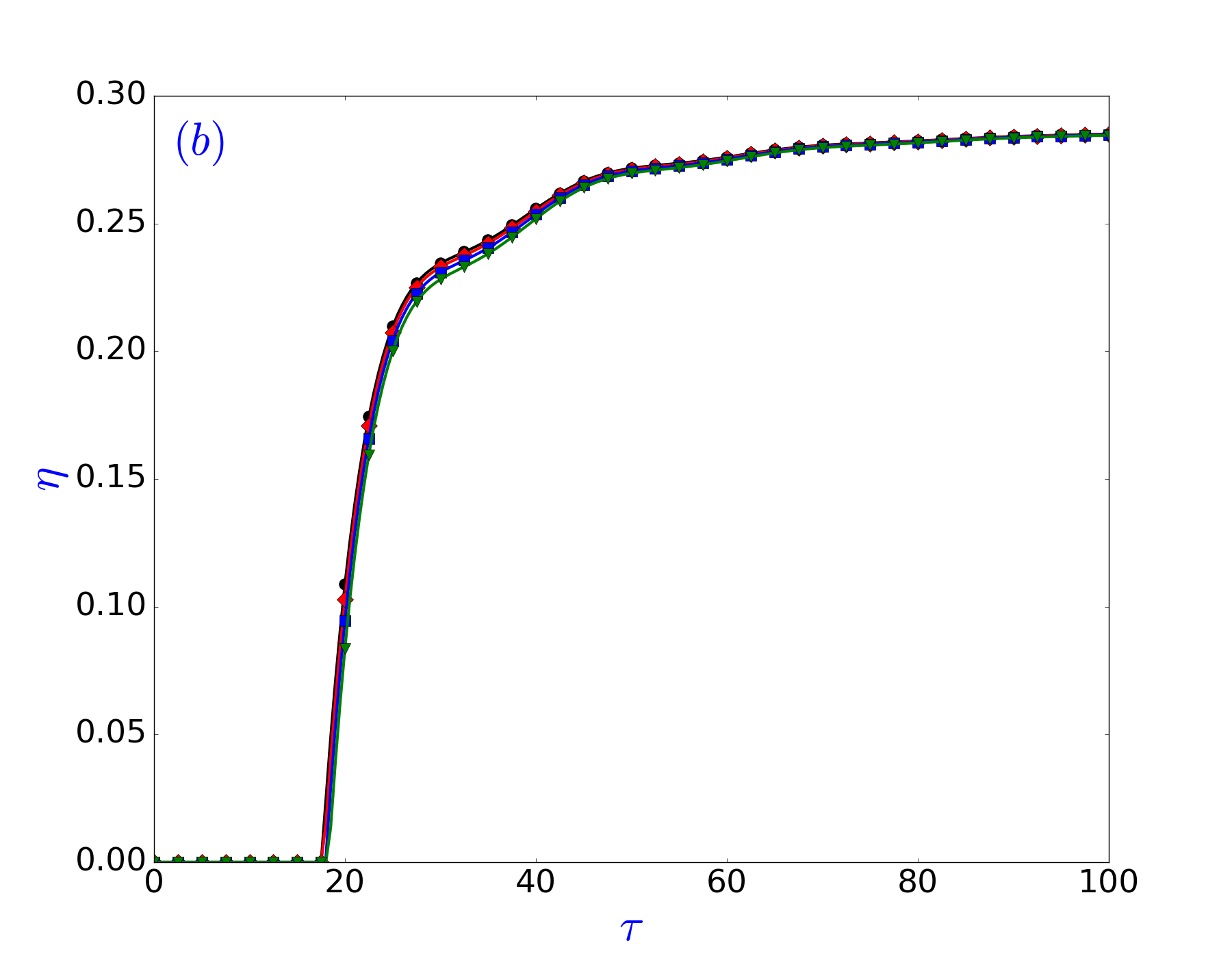}
\caption{\label{fig:1} (Color online.) (a) Work obtained in a cycle and (b) thermal efficiency versus total adiabatic time $\tau$ for the parameters $B_0=B_1=0.5$, $B_2=0.05$, $T_1=2$ and $T_2=1$, and the spins $I=1/2,1,3/2,2$. In this figure, only sinusoidal pulse is considered as a driven pulse to generate the adiabatic branches. Inset in (a) shows the mutual relation between the work output and the total entropy production.}
\end{center}
\end{figure}

Now we generalize the above ideas to an arbitrary spin-$I$ as the working fluid of the Otto engine and plot in Fig.~\ref{fig:1} the work obtained in a cycle and the thermal efficiency as a function of the total adiabatic time $\tau$ generated only by the sinusoidal pulse. The inset in Fig.~\ref{fig:1}(a) shows the mutual relation between $W$ and $\Delta S_E$ for each spin-$I$. The engine can produce larger absolute work, $|W|$, with respect to higher spin-$I$, since higher spins introduce more and higher energy levels. We have numerically verified that the work output for each spin values lies in the range, $W_{\text{lb}} < W < W_{\text{up}}$, where $W_{\text{lb}}$ and $W_{\text{up}}$ are the lower and the upper bounds of the work obtained in the limits $\tau \rightarrow 0$ and $\tau \rightarrow \infty$, respectively. Since the energy gap between adjacent energy levels for each spin-$I$ is given by $\delta_i=\sqrt{B_0^2+B_i^2}$ ($i=1,2$ for the hot and cold isochores, respectively), the efficiency of the engine in Fig.~\ref{fig:1}(b) is bounded above by $\eta_m$~(Eq.~\eqref{eq:eff}). The type of the working fluid has no prominent effect on the critical adiabatic time $\tau_c$ where $W>0$ beyond, and the thermal efficiency of the cycle; each spin-$I$ just introduces tiny shifts to $\eta$ and $\tau_c$. The mutual dependence between the work output and the total entropy production (inset in Fig.~\ref{fig:1}(a)) verifies our preceding conjecture that $\Delta S_E$ and $W$ have linear reciprocal dependence; the increase in the entropy production leads to decrease in the work output, and vice versa. These two quantities are acting as a quantitative measure of the internal friction in our setting. Specifically, for the regions where there is a large entropy production, we have $W<0$. Finally, we note that the main conclusions obtained for the spin-1/2 in Fig.~\ref{fig:0} remains qualitatively the same for the arbitrary spin case and for the other considered driven pulses.

\subsection{\label{sec:intfric} Internal Friction}
In this section, we examine in detail the concept of internal friction and quantitatively determine it in the cycle~\cite{plastina14}. Let us first consider only one adiabatic stage, started from an initial thermal state $\rho_i=\sum_{n} p_n^{(i)} \ket{\epsilon_n^{(i)}}\bra{\epsilon_n^{(i)}}$ with an initial Hamiltonian $H_i$. The reversible adiabatic (frictionless) transformation, performed ideally in an infinite time ($\tau \rightarrow \infty$), ensures the instantaneous follow of the system Hamiltonian. Therefore, a very slow transformation of the system Hamiltonian from an initial value $H_i$ to a final one $H_f$,  will result in a final equilibrium state in the form $\rho_{\text{a}}=\sum_{n} p_n^{(i)} \ket{\epsilon_n^{(f)}}\bra{\epsilon_n^{(f)}}$, where $\ket{\epsilon_n^{(f)}}$ is the eigenstate of $H_{f}$. On the other hand, making the same transformation in a finite time $\tau$, generally, deviates the system final density matrix $\rho_{\tau}$ from equilibrium (coherence generation in energy frame) and may require an extra amount of work. The internal friction, defined as the work done on the system in an actual process differ from the work done in a quasi-static reversible one, equals to the extra energy done on the system by the driving agent against friction. Shortly, $W_{\text{fric}}=W_{\tau}-W_{\tau\rightarrow\infty}=U_{\tau}-U_{\text{a}}\geq 0$, where $U_{\mu}=Tr[H_f\rho_{\mu}]$ ($\mu=\tau,\text{a}$). Ref.~\cite{plastina14} connected the internal friction directly to the quantum relative entropy, $S(\rho||\sigma)=Tr(\rho \ln\rho-\rho \ln\sigma$), between the relevant density matrices as
\begin{equation}\label{eq:wfric}
W_{\text{fric}}= \beta_{\text{a}}^{-1} S(\rho_{\tau}||\rho_{\text{a}})\geq 0,
\end{equation}
where $\beta_{\text{a}}^{-1}$ is the temperature of the equilibrium density matrix $\rho_{\text{a}}$. The non-negativity of the relative entropy according to Klein's inequality guarantees that $W_{\text{fric}}$ is always non-negative~\cite{plastina14}. If we consider the thermalization of the state $\rho_{\tau}$ to the equilibrium state $\rho_{\text{a}}$ by an additional isochoric stage, $W_{\text{fric}}$ can be then interpreted as the waste energy that has to be removed from the system as a heat to go from $\rho_{\tau}$ to $\rho_{\text{a}}$, i.e.,  $W_{\text{fric}}=-Q$, where $Q=U_{\text{a}}-U_{\tau}$ is the heat exchanged during the thermalization process.

\begin{figure}
\begin{center}
\includegraphics[scale=0.21]{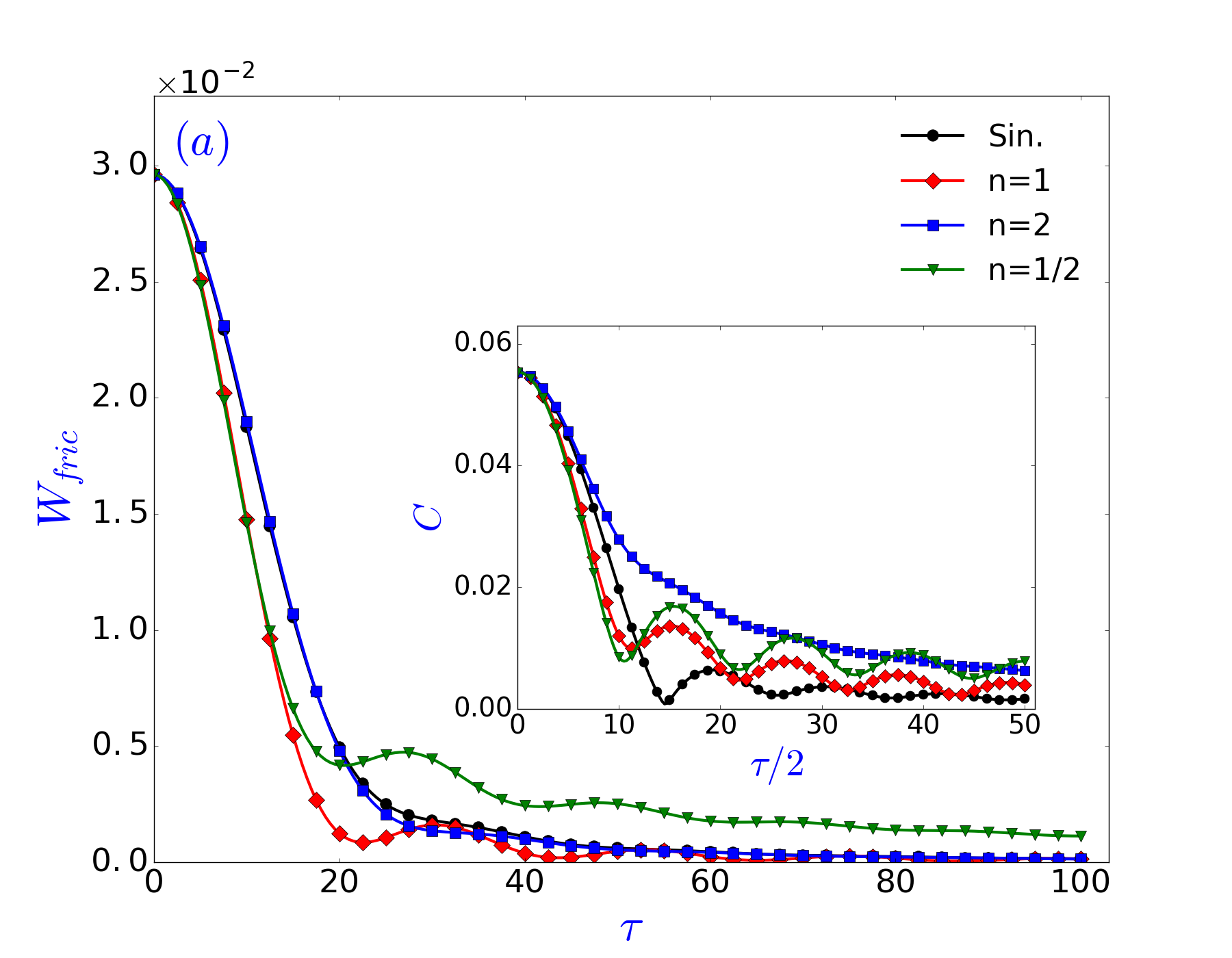}
\includegraphics[scale=0.21]{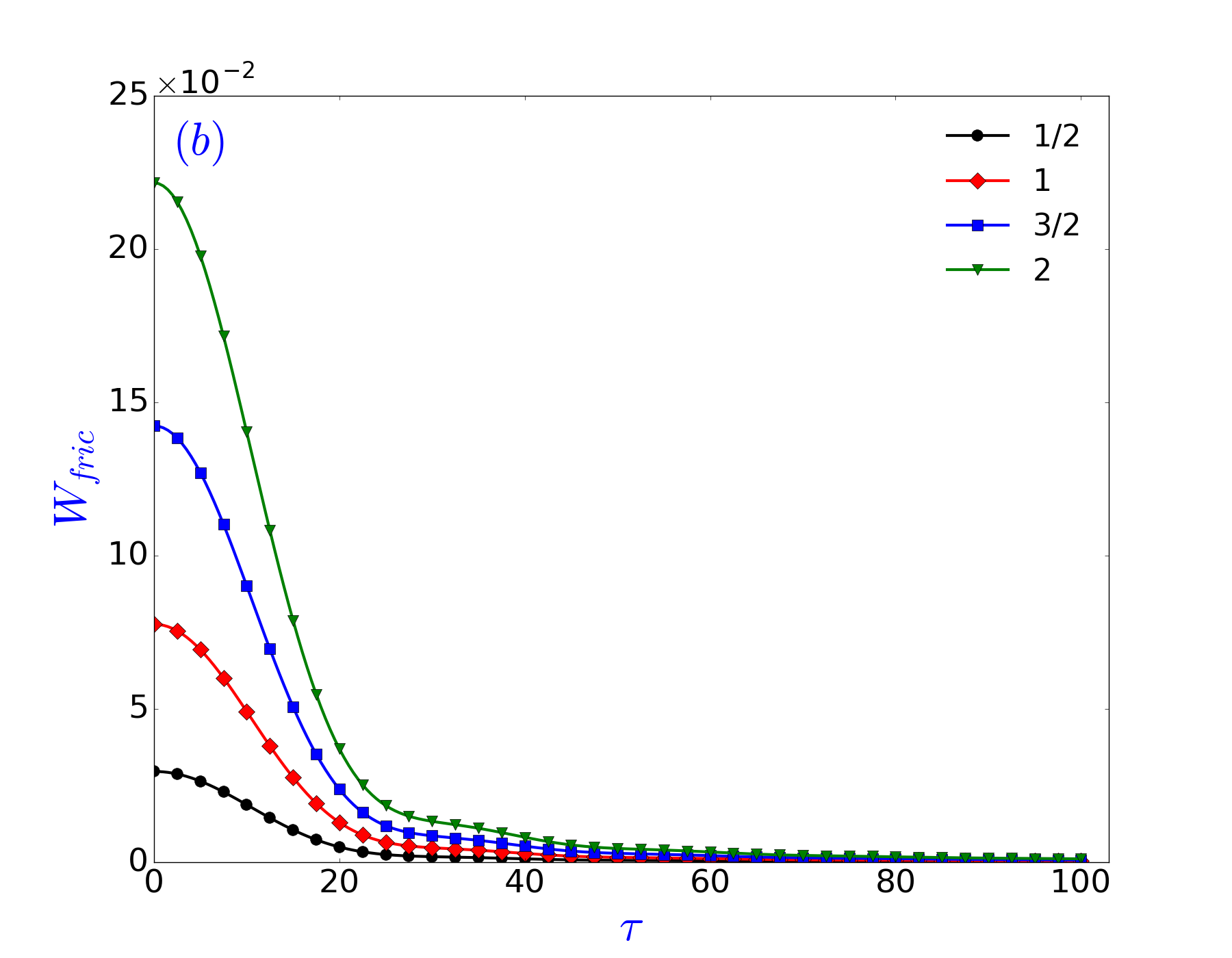}
\caption{\label{fig:2} (Color online.) The total excess energy due to internal friction, i.e., $W_{\text{fric}}$, as a function of total adiabatic time $\tau$ for the parameters and spin values given in (a) Fig.~\ref{fig:0} and (b) Fig.~\ref{fig:1}. The inset in (a) shows an example for the coherence generation in the finite time  process, i.e., $C=\left| \left\langle \epsilon_1^{(f)} \right|\rho_{\tau} \left| \epsilon_2^{(f)} \right\rangle \right|$, as a function of $\tau$ for the spin-1/2 in the adiabatic expansion stroke.}
\end{center}
\end{figure}

By finding the target states $\rho_{\text{a}}$ for the two adiabatic stages and using Eq.~(\ref{eq:wfric}), we determine the internal friction in our Otto cycle and plot it in Fig.~\ref{fig:2} as a function of the total adiabatic time for the spins and the parameters given in Fig.~\ref{fig:0} and Fig~\ref{fig:1}. As expected, $W_{\text{fric}}$ is large for a fast transformation, while it goes to zero only under quantum adiabatic theorem. In fact, the analysis in Fig.~\ref{fig:2} precisely justify our preceding results in Sec.~\ref{sec:workeff}. Here, we explicitly reveal the non-monotone dependence of internal friction on $\tau$ in Fig.~\ref{fig:2}(a) and justify that higher spins are subject to more internal friction in Fig.~\ref{fig:2}(b). In fact, $W_{\text{fric}}$, shown in Fig.~\ref{fig:2}, is exactly the total excess (waste) energy that the system has taken during two adiabatic branches which is then dissipated to the heat baths in the subsequent isochoric stages. This effect, being solely quantum in origin, limits the performance of the quantum Otto engine. If we put our interpretation in a different way, $W_{\text{fric}}$ can be considered as the indicator of the coherence arising in the energy frame in the finite time process. The coherence generation, for instance in the adiabatic expansion stroke for the spin-1/2, is shown in the inset of Fig.~\ref{fig:2}(a). The coherence is erased in the isochoric stages of the cycle as an excess heat as quantified by $W_{\text{fric}}$. The erased information-coherence as a heat may find correspondence through the interpretation of Landauer's principle~\cite{landauer}. We should stress here that the excess work, being quantum or classical in origin, for isolated quantum systems and its non-monotonic feature as a function of switching time have also been discussed in different contents~\cite{cakmak16,acconcia15,acconcia215}.

\section{\label{sec:disc} Discussion: Possible Realization in NMR Systems}
We may suggest that the above predictions can be realized in NMR systems, as they provide successful control on the preparation and dynamics of the systems~\cite{batalhao14,raitz15,oliveira07,stolze08}. Indeed, the non-equilibrium work statistics of $^{13}C$ molecule (spin-1/2) undergoing a closed quasi-unitary evolution is given recently in liquid-state NMR setup~\cite{batalhao14}. Here, the nuclear spin-$1/2$ is driven by a time-modulated radio frequency field with a type of Hamiltonian in the form of Eq.~\eqref{eq:ssh}. 

On the other hand, conventional NMR systems provide a huge magnetic field aligned along one-axis, generally chosen in $z$-axis, comparing with the magnitude of the modulated magnetic field applied on the $x$-$y$ plane. The ratio is generally in the order, $B_{xy}/B_{z} \backsim 10^{-3}-10^{-6}$, where $B_i$ represents the magnetic field strength along plane $i$. We may suggest that under the considered scheme, there would be not possible to device an efficient heat engine, as the maximum efficiency given in Eq.~\eqref{eq:eff} will close to zero. On the other hand, NMR systems also offer the time-dependent changes in the magnetic field along the $z$-axis, rather than the one in $x$-axis. Such techniques are widely used in magnetic resonance imaging to obtain position information. Under this control scheme, the maximum efficiency can be approximated to $\eta \thickapprox 1-B_2/B_1$, where $B_i$ $(i=1,2)$ are the strong magnetic fields in $z$-axis in the hot and cold isochores, respectively. Since there would be weak misalignment due to the weak transverse magnetic field, one may expect weak frictional effects. The numerical simulations of these intriguing directions can be further studied, but goes beyond the aim of the present study.      

\section{\label{sec:concls} Conclusions}
We investigate the performance of a quantum Otto cycle based on a simple experimentally accessible quantum system in the presence of internal friction. The role of total allocated time to the adiabatic branches, generated by different control field profiles, on the extractable work and the thermal efficiency are analyzed in detail. The irreversible nature of the internal friction is characterized by the entropy production and is quantitatively determined by studying the closeness of an actual unitary process to an infinitely long one via quantum relative entropy. The inevitable induced excitations (corresponding to the coherence generation in the eigen-base representation) due to the non-ideal, finite-time adiabatic transformations, when the controlled and free evolutions are not compatible (corresponding terms in the Hamiltonian are not commuting), are found to negatively effect the harvested work and the thermal efficiency of the Otto cycle. The corresponding performance parameters of the cycle are found to strongly depend on the type of the driving magnetic fields on the adiabatic branches. Specifically, each driving pulse are found to define different critical times ($\tau_c$), in which the system acts as a quantum heat engine beyond this critical value. We elucidate the non-monotone character of the work output, thermal efficiency, energy entropy and internal friction as a function of total adiabatic time and the possibility for almost frictionless transformations with small entropy production in finite switching times. Our results show that the proposed model is a promising candidate for efficient and powered QHE implementations in liquid-state NMR setups.

%%%%%%%%%%%%%%%%%%%%%%%%%%%%%%%%%%%%%%%%%%%%%%%%%
\acknowledgments
%%%%%%%%%%%%%%%%%%%%%%%%%%%%%%%%%%%%%%%%%%%%%%%%%
The authors thank Dieter Suter, Sebastian Deffner, Yuanjian Zheng and Avijit Misra for fruitful discussions. S.\c{C}. warmly thanks A. Gen\c{c}ten for useful discussions. F.A. thanks department of Physics of the Ko\c{c} University for hospitality. \"{O}.~E.~M gratefully acknowledges support
by Ko\c{c} University (KU) Visiting Scholar Program by KU Office of VPAA (Vice President of Academic Affairs) and warm hospitality of Quantum Technology Group (QTEQ) at Queen's University Belfast. The authors acknowledge support from University Research Agreement between Ko\c{c} University and Lockheed Martin Corporation.


\begin{thebibliography}{}

\bibitem{scovil59} H.E.D. Scovil and E.O. Schulz-Dubois, Phys. Rev. Lett. \textbf{2}, 262 (1959).

\bibitem{kieu04} T.D. Kieu, Phys. Rev. Lett. \textbf{93}, 140403 (2004).

\bibitem{kieu06} T.D. Kieu, Eur. Phys. J. D \textbf{39}, 115-128 (2006).

\bibitem{thomas11}  G. Thomas and R.S. Johal, Phys. Rev. E \textbf{83}, 031135 (2011).

\bibitem{altintas14} F. Altintas, A.\"{U}.C. Hardal, and \"{O}.E. M\"{u}stecapl{\i}o\u{g}lu, Phys. Rev. E \textbf{90}, 032102 (2014).

\bibitem{quan05} H.T. Quan, P. Zhang, and C.P. Sun, Phys. Rev. E \textbf{72}, 056110 (2005). 	

\bibitem{henrich07} M.J. Henrich, G. Mahler, and M. Michel, Phys. Rev. E \textbf{75}, 051118 (2007).

\bibitem{zhang07} T. Zhang, W.-T. Liu, P.-X. Chen, and C.-Z. Li, Phys. Rev. A \textbf{75}, 062102 (2007).

\bibitem{quan07} H.T. Quan, Y.-X Liu, C.P. Sun, and F. Nori, Phys. Rev. E \textbf{76}, 031105 (2007). 

\bibitem{quan09} H.T. Quan, Phys. Rev. E \textbf{79}, 041129 (2009). 

\bibitem{tonner05} F. Tonner and G. Mahler, Phys. Rev. E \textbf{72}, 066118 (2005).

\bibitem{turkpence16} D. Turkpence and \"{O}.E. M\"{u}stecapl{\i}o\u{g}lu, Phys. Rev. E \textbf{93}, 012145 (2016).

\bibitem{dillen09} R. Dillenschneider and E. Lutz, EPL \textbf{88}, 50003 (2009). 

\bibitem{scully03} M.O. Scully, M.S. Zubairy, G.S. Agarwal, and H. Walther, Science \textbf{299}, 862 (2003). 

\bibitem{hardal15} A.\"{U}.C. Hardal and \"{O}.E. M\"{u}stecapl{\i}o\u{g}lu, Sci. Rep. \textbf{5}, 12953 (2015).

\bibitem{gardas15} B. Gardas and S. Deffner, Phys. Rev. E \textbf{92}, 042126 (2015).

\bibitem{lutz14} J. Robnagel, O. Abah, F. Schmidt-Kaler, K. Singer, and E. Lutz, Phys. Rev. Lett. \textbf{112}, 030602 (2014).	

\bibitem{lutz12} O. Abah, J. Robnagel, G. Jacob, S. Deffner, F. Schmidt-Kaler, K. Singer, and E. Lutz, Phys. Rev. Lett. \textbf{109}, 203006 (2012).

\bibitem{fialko12} O. Fialko and D.W. Hallwood, Phys. Rev. Lett. \textbf{108}, 085303 (2012).	

\bibitem{zhang14} K. Zhang, F. Bariani, and P. Meystre, Phys. Rev. Lett. \textbf{112}, 150602 (2014).	

\bibitem{sothmann12} B. Sothmann, M. B\"{u}ttiker, EPL \textbf{99}, 27001 (2012).

\bibitem{quan06} H.T. Quan, P. Zhang, and C.P. Sun, Phys. Rev. E \textbf{73}, 036122 (2006).	

\bibitem{altintas15} F. Altintas, A.\"{U}.C. Hardal, and \"{O}.E. M\"{u}stecapl{\i}o\u{g}lu, Phys. Rev. A \textbf{91}, 023816 (2015).	

\bibitem{geva92} E. Geva and R. Kosloff, J. Chem. Phys. \textbf{97}, 6 (1992).

\bibitem{rezek10} Y. Rezek, Entropy \textbf{12}, 1885$-$1901 (2010).

\bibitem{kosloff13} R. Kosloff, Entropy \textbf{15}, 2100$-$2128  (2013).

\bibitem{rezek09} Y. Rezek, P. Salamon, K.H. Hoffmann, and R. Kosloff, EPL \textbf{85}, 30008 (2009).

\bibitem{thomas14} G. Thomas and R.S. Johal, Eur. Phys. J. B \textbf{87}, 166 (2014).

\bibitem{campisi15} M. Campisi, J. Pekola, and R. Fazio, New J. Phys. \textbf{17}, 035012 (2015).

\bibitem{alecce15} A. Alecce, F. Galve, N.L. Gullo, L. Dell'Anna, F. Plastina, and R. Zambrini, New J. Phys. \textbf{17}, 075007 (2015).

\bibitem{rezek06} Y. Rezek and R. Kosloff, New J. Phys. \textbf{8}, 83 (2006).

\bibitem{wang07} J. Wang, J. He, and Y. Xin, Phys. Scr. \textbf{75}, 227$-$234 (2007).

\bibitem{feldmann00} T. Feldmann and R. Kosloff, Phys. Rev. E \textbf{61}, 4774 (2000).

\bibitem{kosloff02} R. Kosloff and T. Feldmann, Phys. Rev. E \textbf{65}, 055102(R) (2002). 

\bibitem{feldmann03} T. Feldmann and R. Kosloff, Phys. Rev. E \textbf{68}, 016101 (2003). 

\bibitem{feldmann04} T. Feldmann and R. Kosloff, Phys. Rev. E \textbf{70}, 046110 (2004). 
 
\bibitem{allahverdyan05} A.E. Allahverdyan and T.M. Nieuwenhuizen, Phys. Rev. E \textbf{71}, 046107 (2005). 

\bibitem{feldmann06} T. Feldmann and R. Kosloff, Phys. Rev. E \textbf{73}, 025107(R) (2006). 

\bibitem{kosloff10} R. Kosloff and T. Feldmann, Phys. Rev. E \textbf{82}, 011134 (2010).

\bibitem{wang12} J. Wang, J. He, and Z. Wu, Phys. Rev. E \textbf{85}, 031145 (2012).

\bibitem{feldmann12} T. Feldmann and R. Kosloff, Phys. Rev. E \textbf{85}, 051114 (2012).

\bibitem{wang13} R. Wang, J. Wang, J. He, and Y. Ma, Phys. Rev. E \textbf{87}, 042119 (2013).

\bibitem{torron13} E. Torrontegui and R. Kosloff, Phys. Rev. E \textbf{88}, 032103 (2013). 

\bibitem{ahn03} K.-H. Ahn and P. Mohanty, Phys. Rev. Lett. \textbf{90}, 8 (2003).

\bibitem{deffner10} S. Deffner and E. Lutz, Phys. Rev. Lett. \textbf{105}, 170402 (2010). 

\bibitem{wu06} F. Wu, L. Chen, F. Sun, C. Wu, and Q. Li, Phys. Rev. E \textbf{73}, 016103 (2006).

\bibitem{plastina14} F. Plastina, A. Alecce, T.J.G. Apollaro, G. Falcone, G. Francica, F. Galve, N.L. Gullo, and R. Zambrini, Phys. Rev. Lett. \textbf{113}, 260601 (2014).
  
\bibitem{campo15} A. del Campo, J. Goold, and M. Paternostro, Sci. Rep. \textbf{4}, 6208 (2014).

\bibitem{zheng15} Y. Zheng, S. Campbell, G.D. Chiara, and D. Poletti, arXiv:1509.01882. 

\bibitem{zheng16} Y. Zheng, P. Hanggi and D. Poletti, arXiv:1604.00489.

\bibitem{Ribeiro16} W. L. Ribeiro, G. T. Landi, F. L. Semiao, arXiv:1601.01833. 

\bibitem{misra15} A. Misra, U. Singh, M.N. Bera and A.K. Rajagopal, Phys. Rev. E \textbf{92}, 042161 (2015).

\bibitem{batalhao14} T.B. Batalhao, A.M. Souza, L. Mazzola, R. Auccaise, R.S. Sarthour, I.S. Oliveira, J. Goold, G.D. Chiara, M. Paternostro, and R.M. Serra, Phys. Rev. Lett. \textbf{113}, 140601 (2014). 

\bibitem{raitz15} C. Raitz, A.M. Souza, R. Auccaise, R.S. Sarthour, and I.S. Oliveira, Quantum Inf. Process. \textbf{14}, 37-46 (2015).

\bibitem{cakmak16} S. Cakmak, F. Altintas and O.E. Mustecaplioglu, Phys. Scr. \textbf{91}, 075101 (2016).

\bibitem{acconcia15} T.V. Acconcia, M.V.S. Bonanca and S. Deffner, Phys. Rev. E \textbf{92}, 042148 (2015).

\bibitem{acconcia215} T.V. Acconcia and M.V.S. Bonanca, Phys. Rev. E \textbf{91}, 042141 (2015).

\bibitem{oliveira07} I.S. Oliveira, T.J. Bonagamba, R.S. Sarthour, J.C.C Freitas, and E.R. deAzevedo, \textit{NMR Quantum Information Processing} (Elsevier, 2007).

\bibitem{stolze08} J. Stolze and D. Suter, \textit{Quantum Computing: A Short Course from Theory to Experiment} (Wiley, 2008).

\bibitem{landauer} R. Landauer, IBM J. Res. Develop. \textbf{5}, 183-191 (1961).


\end{thebibliography}
\end{document}